\documentclass[fleqn,10pt]{wlscirep}
\usepackage{csquotes}
\usepackage[utf8]{inputenc}
\usepackage{newunicodechar}
\newunicodechar{−}{\ensuremath{-}}
\usepackage[T1]{fontenc}
\usepackage{amsmath}
\usepackage{amsfonts}
\usepackage{amssymb}
\usepackage{comment}
\usepackage{graphicx}
\usepackage{booktabs, longtable, makecell, multirow, xltabular}
\usepackage{xr}
\usepackage{soul}

\usepackage[sorting=none]{biblatex} 
\bibliography{bibliography}

\newcommand{\norm}[1]{\left\lVert#1\right\rVert}
\title{Practical protein-pocket hydration-site prediction for drug discovery on a quantum computer}

\author[1,+]{Daniele Loco}
\author[2,+]{Kisa Barkemeyer}
\author[2]{Andre R. R. Carvalho}
\author[1,3,*]{Jean-Philip Piquemal}
\affil[1]{Qubit Pharmaceuticals, Advanced Research Team, 75014 Paris, France}
\affil[2]{Q-CTRL, Berlin, Germany and Sydney, NSW Australia}
\affil[3]{Sorbonne Université, LCT, UMR 7616 CNRS, 75005 Paris, France}

\affil[*]{Email: jean-philip.piquemal@sorbonne-universite.fr}

\affil[+]{these authors contributed equally to this work}

\begin{abstract}
Demonstrating the practical utility of Noisy Intermediate-Scale Quantum (NISQ) hardware for recurrent tasks in Computer-Aided Drug Discovery is of paramount importance. We tackle this challenge by performing three-dimensional protein pockets hydration-site prediction on a quantum computer. Formulating the water placement problem as a Quadratic Unconstrained Binary Optimization (QUBO), we use a hybrid approach coupling a classical three-dimensional reference-interaction site model (3D-RISM) to an efficient quantum optimization solver, to run various hardware experiments up to 123 qubits. Matching the precision of classical approaches, our results reproduced experimental predictions on real-life protein-ligand complexes. Furthermore, through a detailed resource estimation analysis, we show that accuracy can be systematically improved with increasing number of qubits, indicating that full quantum utility is in reach as devices scale in the near term. 
The method has potential for assisting simulations of protein-ligand complexes for drug lead optimization and setup of docking calculations. 
\end{abstract}

\makeatletter
\newcommand*{\addFileDependency}[1]{
  \typeout{(#1)}
  \@addtofilelist{#1}
  \IfFileExists{#1}{}{\typeout{No file #1.}}
}
\makeatother

\newcommand*{\myexternaldocument}[1]{%
    \externaldocument{#1}%
    \addFileDependency{#1.tex}%
    \addFileDependency{#1.aux}%
}
\myexternaldocument{SM}

\begin{document}

\flushbottom
\maketitle
%
%

\section*{Introduction}

In the past 20 years, Computer-Aided Drug Discovery (CADD) has become an important tool for the pharmaceutical industry to enhance drug discovery  pipelines~\cite{CADD_RV_1,CADD_RV_2,CADD_RV_3}. Accurate numerical methods capable of modeling chemical processes, physical interactions between complex molecular structures, and dynamical effects occurring in biological systems are of great value, improving experimental data interpretability, while potentially reducing the overall costs thanks to improved theoretical predictions.

Between the many challenges faced in the long and expensive process of developing a drug, there is the need to understand the role of water. Water, as a primary component of biological environments, profoundly influences the overall structural stability of proteins, as well as the interactions of proteins with small molecules, often referred to as ligands,  making the understanding of its behavior fundamental for estimating drug binding and optimizing drug candidates~\cite{WP1,Blazhynska2025.10.17.683050,WM1}. Water can act as an intermediate between a protein and a ligand, called interfacial waters, playing a fundamental role in the binding process~\cite{WP1,WB_3}. However, predicting the configuration of water molecules presents a formidable computational challenge due to the high dimensionality of the problem and the necessity to sample a vast conformational space~\cite{Blazhynska2025.10.17.683050,WM0,WM1,WM2,WM3}. 

Classically, molecular dynamics and Monte Carlo strategies are widely used to address the problem of configurational sampling in an effective way~\cite{WM0,lambdaabf,lambdaabfopes,jorgensenMC,dmccammonreview,Hénin_Lelièvre_Shirts_Valsson_Delemotte_2022}. These types of algorithms are general enough to handle the configurational sampling of molecules larger than water, such as biomolecules and large aggregates. However, sampling water will always remain difficult due to the complexity of the networks of hydrogen bonds it can form, while either exchanging rapidly with bulk solvent or, conversely, becoming structurally stabilized in binding sites~\cite{bodnarchuk2014strategies,de2010role,setiadi2025thermodynamics,ge2022enhancing}. Modern enhanced sampling techniques enable to precisely capture the water-protein and water-ligand interactions in complex biological systems~\cite{D1SC00145K,interfacewater,schahl2025histidine} to provide predictive drug binding free-energy estimation~\cite{Blazhynska2025.10.17.683050,D1SC05892D,ansari2025targeting}. This, however, comes with a significant computational cost and state-of-the-art simulation codes heavily rely on massive parallelism and GPU-acceleration~\cite{C7SC04531J,adjouatinkerhp,kohnke2020gpu,phillips2020scalable,ambergpus}.  
In parallel, promising machine learning approaches~\cite{bigi2025flashmdlongstrideuniversalprediction, plainer2025consistentsamplingsimulationmolecular} have been steadily improving and becoming able to sample molecular equilibrium structures. However, to date, these approaches do not yet reach sufficient accuracy for CADD applications ~\cite{annurevnoe,CADD_RV_3, QAB_1}. 

Alternatively, quantum computing holds the promise of addressing these classically hard computational tasks. Indeed, quantum speedups are expected to be substantial for implementations of statistical physics problems related to configuration sampling~\cite{Markov25,Metropolis25}. A number of studies have presented applications of different types of quantum algorithms to biomolecular systems~\cite{robert_resource-efficient_2021, QAB_1, liliopoulos_quantum_2025, Ding_molecular_docking_2024} and, more generally, full drug discovery pipelines that incorporate quantum-based computations have been proposed and reviewed in the literature~\cite{QC_DD_1,QC_DD_2}.

In this context, some of us recently presented an approach for the quantum prediction of protein pockets hydration~\cite{QP}. Using a Quadratic Unconstrained Binary Optimization (QUBO) cost function embodying the water placement problem, we mapped the problem of identifying hydration sites onto an Ising Hamiltonian. This mapping enabled the implementation of an adiabatic protocol on an analog quantum device based on Rydberg cold atom physics~\cite{QP}. The method used 2D slices of 3D-RISM (Reference Interaction Site Model) continuous water density functions to define a discrete optimization problem, which was then mapped onto the Hamiltonian of the corresponding 2D Rydberg atoms slice\cite{QP}. Although this work relied on 2D slices only, which poses challenges for applicability to 3D systems, it represented a key step toward the use of quantum computing in CADD. 

In this paper, taking further steps towards developing practical quantum approaches for CADD, we present an extension of the previous work~\cite{QP}, leveraging digital quantum computing to predict the 3D hydration sites in protein pockets in real-life conditions using quantum hardware. Our methodology integrates a classical formulation of the hydration-site prediction problem with quantum optimization techniques, simultaneously leveraging advanced software~\cite{ES_1, ES_2} and hardware to address the problem on a quantum computer. The presented end-to-end workflow includes the problem encoding, the quantum algorithm implementation, and hardware execution using an efficient quantum solver~\cite{ES_2}, enabling us to explore the practical utility for large-scale instances of the problem on current Noisy Intermediate-Scale Quantum (NISQ) devices. This strategy allowed us to describe realistic 3D systems without additional approximations~\cite{QP}. We applied our approach to a selected list of proteins relevant to the pharmaceutical industry, encompassing FDA (U.S. Food and Drug Administration) approved drugs in complex with their target protein. Our results were obtained from experiments performed on the latest Heron 156-qubits IBM devices and compared to existing classical solvers. Finally, we discuss the pathway to quantum advantage by presenting a resource estimation analysis to forecast the device specifications needed to address protein complexes at scales where classical methods struggle to find solutions.

\section*{Results}

Our hydration-sites prediction workflow consists of multiple steps that are described in detail in the Methods section and summarized in Figure~\ref{fig:flow}. It starts from the calculation of 3D-RISM densities ($g(\mathbf{r})$) from the 3D molecular structure of the target protein for the hydration site prediction. This 3D-RISM density contains the essential information about the water distribution around the protein. We then describe this density as a sum of Gaussian distributions whose centers are associated to the location of the water molecules and with variances corresponding to the uncertainty in the water position. Therefore the problem of locating the positions of water molecules is mapped onto the problem of finding the best Gaussian mixture that approximates the 3D-RISM density. A key element put forward in~\cite{QP} and further explored here is framing this as a QUBO problem amenable to a quantum optimization. Once this optimization is performed, the outputs are the 3D spatial coordinates of the most likely sites for stable water molecules.

This Results section is divided into five subsections reflecting the workflow described above. We first present the set of protein complexes selected for this study, including the relevant discretization and Gaussian parameters used to define our test cases. We then report on the results of the crucial optimization step executed on quantum computers, up to the current hardware limits. To understand the performance of the method beyond this limit, we also present results for large scale systems using classical approximate optimization approaches. Finally we connect these optimization results to the other components of the workflow to obtain our final hydration sites prediction. These results are on par with, or sometimes better, than other hydration prediction methods available in the literature. At the end of the section we forecast the resources needed on a quantum device to achieve such problem scales.

\subsection*{Test systems: realistic drug-protein complexes}

We start by choosing the protein 3D structures required in the first step of our workflow, shown in Figure~\ref{fig:flow} in Methods. We tested our approach on a set of proteins of interest for health care applications, all of them being related to human diseases for which a drug has been already developed, and later approved by the FDA agency~\cite{R1_FDA}. Table~\ref{tab:test_set} shows the crystal structures used in this work, each identified by their Protein Data Bank~\footnote{https://www.rcsb.org/} (PDB) ID. We focus our attention only on the binding site of each protein, and show in Table~\ref{tab:test_set} the number of crystal water (CW) molecules in the crystallographic structure inside the binding pocket, the number of protein residues identifying the pockets, the protein molecular weight (MW), and the resolution of the x-ray diffraction experiments used to determine the molecular structure.
 
The first five structures were selected out of the 108 analyzed by Samways et al. in~\cite{R1_FDA}, and they all include protein structures with a ligand bound to it, which are commonly denoted as protein-ligand complexes, or holo structures. Our selection was kept small enough to be compatible with the quantum computing resources available for our experiments, while including proteins of different sizes and families to guarantee sufficient diversity. For these structures, we can decide if we include or remove the explicit ligand molecular structure for the hydration-site prediction. These will correspond to two different predictions, as we will show later. Additionally, we included in this list the example of the apo structure of PDB ID 3b7e (PDB ID $=$ 3beq). The term apo refers to the protein 3D conformation where no ligand is bound to it. Knowing the differences in the protein hydration between holo and apo structures is an important information for drug discovery modelers. In practice, knowing the hydration in the apo state, can help the design of new efficient binders, mimicking  the protein-water interactions observed in the corresponding holo structure. This additional example complements our test set to further show the potential of our approach for applications to real use-cases in drug discovery.

\begin{table}[!htb]
\begin{tabular}{|c|c|c|c|c|}
\toprule
PDB ID & \text{\#}  CW & \# residues & MW (kDa)& Resolution ($\text{\AA}$)\\
\midrule
    1f9g  & 6  & 25 &83.7 & 2.0\\
    1x70  & 16 & 18 &174.9 & 2.1\\
    2f9w  & 15 & 10 &59.9  & 1.9\\
    3b7e  & 28 & 15 &86.8  & 1.45\\
    4h2f  & 13 & 27 &60.8  & 1.85\\
    3beq  & 10 & 15 &87.0  & 1.64\\
\bottomrule
\end{tabular}
\caption{Selection of protein structures considered in this study. The columns represent the protein PDB ID, the number of crystal water (CW) molecules present in the crystallographic structure inside the binding pocket, the number of protein residues (amino acids) that identify the binding pockets, the protein molecular weight (MW), and the resolution of the x-ray diffraction experiment that determined the molecular structure. PDB IDs 3b7e and 3beq are, respectively, the IDs used to register a holo and a apo structure of the same protein.}
\label{tab:test_set}
\end{table}

\subsubsection*{Dimension reduction}

Mapping the full 3D-RISM densities of the large selected systems would result in QUBO instances of $\sim$1000000 binary variables, which would make the corresponding optimization problem unattainable for both classical methods and current quantum computers. 
Therefore, to simplify the problem, one needs to define the problem on a grid coarser than that of the initial high-resolution 3D-RISM density. For this, we re-sample $g(\mathbf{r})$ using a grid spacing $\delta$ larger than the original 3D-RISM one ($\delta = 0.5 \textbf{\AA}$), and remove low density values according to an arbitrary threshold value $\tau_g$ (see Methods). Such a strategy impacts the hydration-site prediction accuracy, as discussed later in text, but enables us to address the problem within a reasonable time using modest classical computing resources, while at the same time 
making the problem suitable for execution on quantum hardware by choosing parameters compatible with current limitations on qubit number and two-qubit gates. 
The QUBO instances resulting from this resampling procedure are shown in Table~\ref{tab:systems}, together with the relevant parameters used to pre-process the computed 3D-RISM density and to map them onto each QUBO problem instance. The structures considered are either a holo structure with ligand (L), a holo structure without ligand (NL), or the apo structure (APO) from Table~\ref{tab:test_set}.

\begin{table}[htb]
\begin{tabular}{|c|c|c|c|c|c|}
\toprule
Instances & $\#$ variables & $\delta $ & $\tau_g$ & $\sigma^2$  & label\\
\midrule
   1f9g (L)  & 98  &    1.05   &  0.17   & 0.8 & a \\
    1f9g (NL) & 91  &   1.05   &  0.17   & 0.8 & b \\
    1x70 (L)  & 69  &   0.95   &  0.10   & 0.8 & c \\
    1x70 (NL) & 116 &   0.95   &  0.10   & 0.8 & d \\
    2f9w (L)  & 41  &   1.05    &  0.10   & 0.8 & e \\
    2f9w (NL) & 92  &   1.05   &  0.10   & 0.8 & f \\
    3b7e (L)  & 71  &   1.15   &  0.10   & 1.0 &    g \\
    3b7e (NL) & 90  &   1.15   &  0.10   & 1.0 &    h \\
    4h2f (L)  & 82  &   0.95   &  0.12   & 0.8 & i \\
    4h2f (NL) & 79  &   0.95   &  0.12   & 0.8 & l \\
    3beq (APO) & 72  &   1.25   &  0.10   & 1.0 & m \\
\bottomrule
\end{tabular}
\caption{Test cases listed by PDB ID and labeled according to the inclusion (L) or exclusion (NL) of the ligand. We show the number of binary variables of the associated QUBO problem, the grid parameters ($\delta$ and $\tau_g$), as well the variance of the Gaussian distributions used for each problem instance. We label each of them with the characters from the last column for easy of reference throughout the text. Additional details about the parameters choice can be found in Section~S2 of the Supplementary Information (SI).}
\label{tab:systems}
\end{table}

\subsection*{Execution of 3D hydration-site prediction on quantum hardware}

\begin{figure}[htb]
    \centering
\includegraphics[width=0.5\linewidth]{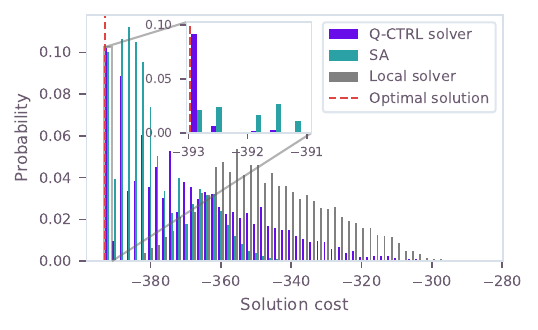}
\caption{Cost distributions for the hydration-site prediction QUBO problem instance d as specified in Table~\ref{tab:systems}, requiring 116 qubits. The Q-CTRL solver output obtained using IBM Kingston (purple bars) is compared to the optimal solution calculated classically with CPLEX (dashed red line). For reference, the results of simulated annealing (SA, teal bars) as well as the results of a greedy local solver (gray bars) are displayed. Inset: Zoomed-in view of the near-optimal region using finer bins.}
    \label{fig:cost_distributions_instance_d}
\end{figure}

To assess the performance of the quantum optimization step for our 3D hydration-site prediction workflow, we solved the underlying QUBO problem for the instances described in Table~\ref{tab:systems} on IBM devices using Q-CTRL's quantum optimization solver (see Methods for details). Exemplary results for the largest instance considered within the test set, instance ``d'' in Table~\ref{tab:systems}, which requires 116 variables corresponding to 116 qubits, are shown in Figure~\ref{fig:cost_distributions_instance_d}. It displays the cost distribution of sampled bitstrings together with the optimal solution calculated classically using CPLEX. For the considered problem instance size, finding provably optimal solutions is still possible, while it becomes increasingly difficult for larger-scale instances as we show below. The Q-CTRL solver finds the optimal solution with a sampling probability of around 9\%, a significantly higher success probability than that of simulated annealing (SA) used as a classical reference heuristic. SA results were obtained with the neal library~\footnote{https://github.com/dwavesystems/dwave-neal} using a geometric temperature schedule and 10,000 samples to evaluate the final cost distribution. Although providing an overall distribution favoring lower cost solutions, SA identifies the optimal solution with a probability of only about 2\%. In contrast, a local solver that applies greedy optimization to randomly sampled candidate solutions is significantly outperformed and fails to identify the optimal solution. Details on the implementation of the local solver which can be interpreted as a zero-temperature version of SA can be found in~\cite{sachdeva_quantum_2024}.

\begin{figure}
    \centering
\includegraphics[width=0.5\linewidth]{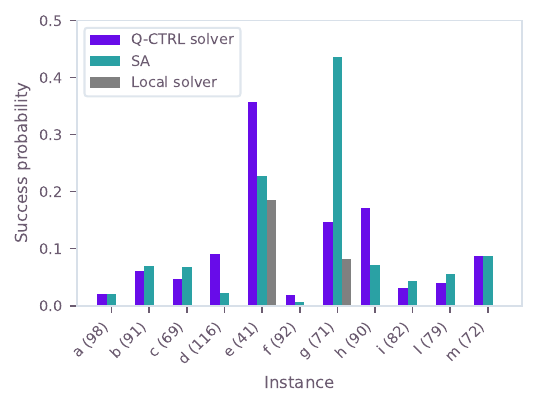}
\caption{Probability to sample the optimal solution using the Q-CTRL solver on IBM Kingston (purple bars), simulated annealing (SA, teal bars), and a greedy local solver (gray bars) for the test-set instances labeled according to Table~\ref{tab:systems}, with variable numbers indicated in parentheses.}
    \label{fig:best_sol_comparison}
\end{figure}

The same process of analyzing the QPU output and comparing it to the exact CPLEX solution was repeated for the whole test-set to extract the probability of obtaining optimal solution. These probabilities are shown in Figure~\ref{fig:best_sol_comparison} for all instances, demonstrating that the quantum solver executed on real hardware is able to successfully solve the hydration-site prediction QUBO problem for a variety of systems with a performance that is overall comparable to SA. Notably, the quantum solver identifies the optimal solution in all cases. By contrast, the greedy local solver only identifies the optimal solution with a non-negligible probability for a few smaller instances and deviates from optimality as instance size grows (see SI, Section~S5 for additional details on the cost gaps). 

Since for the instances considered in this work (Table~\ref{tab:systems}) exact classical solvers find provably exact solutions quickly, in practice, we would not deploy quantum resources for these cases. Nevertheless, as the problem size grows, exact methods typically scale poorly, and beyond a certain point, high-quality heuristics, classical or quantum, become preferable. We, thus, aim to identify a point where the performance of exact classical approaches for solving these problems deteriorates, leaving room for heuristic approaches, such as the Q-CTRL quantum optimization solver, to yield better performance.

\begin{figure}[htb]
    \centering
\includegraphics[width=\linewidth]{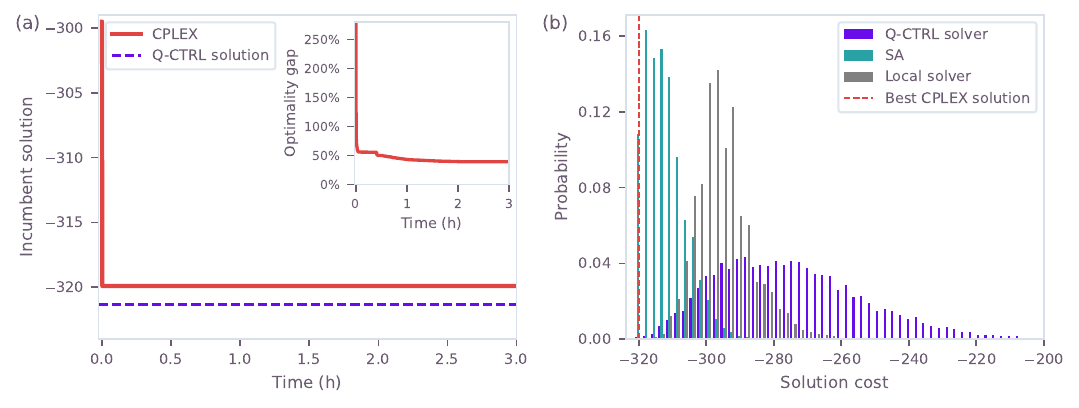}
\caption{Results for the 3b7e-with-ligand instance with a grid spacing of  1.9\,Å corresponding to 123 variables. (a) Exact classical (CPLEX) results. Main: Incumbent solution (red line) over time compared to the best solution identified using the Q-CTRL solver on IBM Pittsburgh (dashed purple line). Inset: Optimality gap over time. (b) Cost distributions for the Q-CTRL solver on IBM Pittsburgh (purple bars), simulated annealing (teal bars), and a greedy local solver (gray bars) shown together with the best CPLEX solution (dashed red line).}
    \label{fig:opt_gap}
\end{figure}

To determine the hardness of the hydration-site prediction QUBO problems, we benchmark how the computational complexity increases with problem scale for the exemplary case of the 3b7e-with-ligand structure. Using CPLEX on a standard laptop, we consider problem instances at a range of scales, obtained by varying the grid spacing. We observe that for 123 variables, corresponding to a grid spacing of 1.9\,Å, it is no longer possible to find the optimal solution within a cut-off time of three hours. As shown in Figure~\ref{fig:opt_gap}-a, the incumbent solution only improves in the first few steps before plateauing. The optimality gap indicates how far the incumbent solution may be from the true optimum and thus acts as a proxy for solution quality. As shown in the inset of Figure~\ref{fig:opt_gap}-a, the optimality gap stalls around 40\%, indicating that, in principle, there is still significant room for improvement of the solution. In fact, the incumbent CPLEX solution after the cut-off time ($C_\text{best, CPLEX} = -319.90$) is not the optimal solution: running the same instance with the Q-CTRL solver yields a lower cost ($C_\text{best, Q-CTRL} = -321.32$) after a total runtime of only 25 minutes on IBM Pittsburgh.

\begin{figure}
    \centering
\includegraphics[width=.5\linewidth]{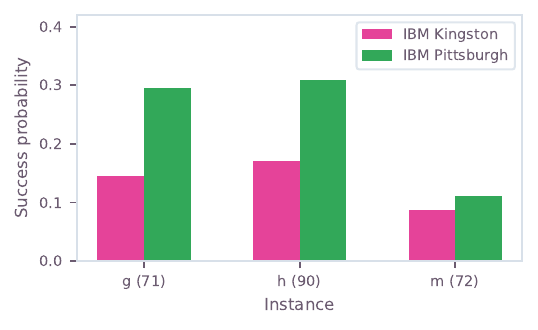}
\caption{
Probability to sample the optimal solution using the Q-CTRL solver on the Heron r2 backend ibm\_kingston (red bars) and the Heron r3 backend ibm\_pittsburgh (blue bars) for a selection of the test-set instances labeled according to Table~\ref{tab:systems}, with variable numbers indicated in parentheses.}
    \label{fig:best_sol_comparison_qpus}
\end{figure}

Note that the results presented in Figure~\ref{fig:opt_gap}-a indicate that the instance is challenging to solve for an exact solver and one needs to resort to approximate methods. As we see in the output cost distributions in Figure~\ref{fig:opt_gap}-b, both the heuristic SA and the Q-CTRL solver yield the same best solution cost, with the former achieving higher probability. This can be attributed to the fact that the quantum resources required to run this instance reach current device limits in terms of gate fidelities, circuit depth, and T1 times.
To reliably tackle even larger problems, quantum algorithm development and further quantum hardware improvements towards larger and better devices must go hand in hand. The effect of the progress on the hardware side was evident during our study: while we were producing our results, IBM’s newest Heron r3 processor, the ibm\_pittsburgh backend, came online. We ran a subset of test-set instances on both Heron r2 and r3 devices and observed significant performance differences, as shown in Figure~\ref{fig:best_sol_comparison_qpus}. This highlights the potential for substantial improvements, including the possibility of outperforming SA, as next-generation devices arrive in line with IBM’s quantum development roadmap \cite{noauthor_ibm_nodate}. Expected future developments are discussed in more detail in the Advantage forecasting section.

\subsection*{Performance evaluation of 3D hydration-site prediction}
\label{subsec_evaluationpredictions}

Now that we have shown that, for the systems considered so far, the optimization step of our workflow can be successfully solved using Q-CTRL's solver on a quantum computer, we can look at how these results impact the ultimate goal of predicting the hydration sites. 

The first step is to consider how to interpret the output bitstrings corresponding to the optimal solutions as spatial water configurations. In our QUBO formulation (see Methods), each qubit has been mapped onto a spatial location, and the values $0$ and $1$ correspond, respectively, to the absence or presence of a water molecule at the point in space associated to that register. Thus we read the corresponding 3D coordinates associated to qubits in state 1 and place a water at that point in space. In practice we only place the oxygen atom of the water molecule, which is sufficient to identify its location. Repeating this operation for each element of a bitstring, we obtain the predicted water (PW) molecules configuration. 

The next step is to assess the accuracy of these predictions. It is common practice in the field to use the crystallographic water (CW) positions to assess the quality of a given numerical method aiming at identifying hydration sites~\cite{WM0,WM1,WM2,WM3,WM4,jorgensenMC}. Here, we will use the CWs available in the experimental data as a reference. One metric we use to asses the quality of our predictions is the percentage of CWs that we are able to identify. We use the character C to refer to this metric, which is computed searching for the presence of at least one PW in a radius of 3 $\textbf{\AA}$ around each CW. If more than one PW is found, we also record how many additional PWs are associated to each CW. We consider such set of PWs as belonging to the same cluster. Performing this analysis for each CW, we use as an additional evaluation metric the average number of PWs found per CW, and call it the average cluster size ($\langle \mathrm{CS} \rangle$). We are also interested in determining how spatially close the PWs are from the CWs. We denote this third metric the precision (P), that we measure both for the closest PW associated to each CW, namely P$^*$, and for the ensemble of the cluster associated to each CW ($\langle \mathrm{P} \rangle$). The definitions of these metrics are reported in the Methods section.

For this performance evaluation, different aspects need to be considered. In the first place, it is important to note that the crystal structures used in this work have been determined with a reasonably good experimental resolution (see Table~\ref{tab:test_set}). This ensures that the comparison between PW and CW is meaningful. 
Furthermore, the accuracy of our QUBO-based approach also depends on the underlying 3D-RISM densities and on how well they actually represent the water molecules distribution in the protein. Available benchmarks from the literature assess this factor, proving a generally good cost-balanced 3D-RISM's quality for hydration free energy prediction~\cite{R2_2_3DRISM2}, in line with the quality-requirements of fast virtual screening tasks and systems preparation in CADD. On top of these factors, the discretization step and dimensionality reduction procedure we perform to map the 3D-RISM density into QUBO (see Figure~\ref{fig:flow}) also impact the method's accuracy. In particular we can expect to obtain the best mapping between the continuous density and the QUBO when $\delta$ is equal or smaller than the spacing of the $g(\mathbf{r})$ grid (0.5 $\text{\AA}$) and if the $\tau_g$ threshold remains negligibly small. 

\begin{figure}[!htb]
    \centering
\includegraphics[width=.87\textwidth]{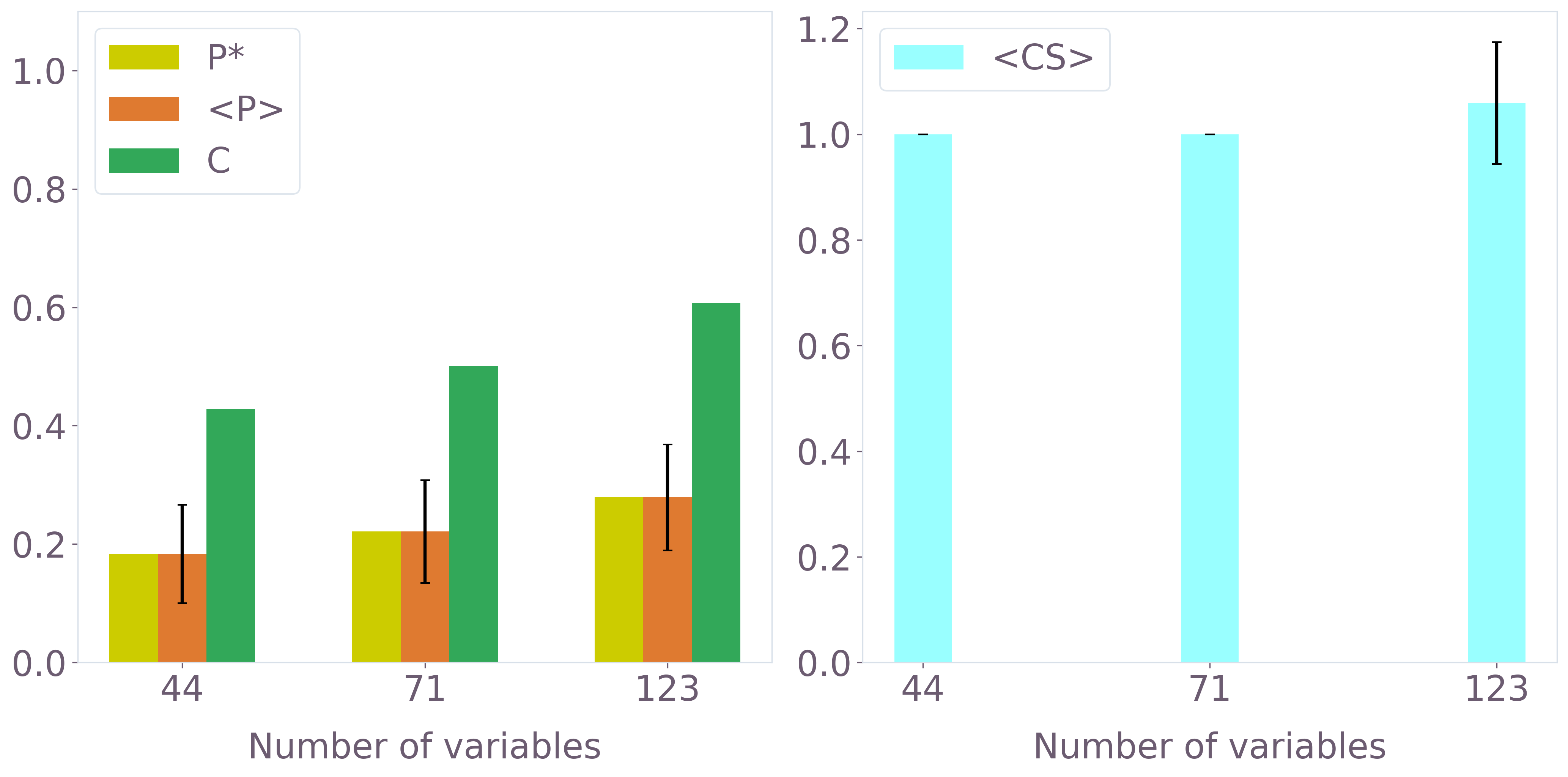}
\includegraphics[width=.87\textwidth]{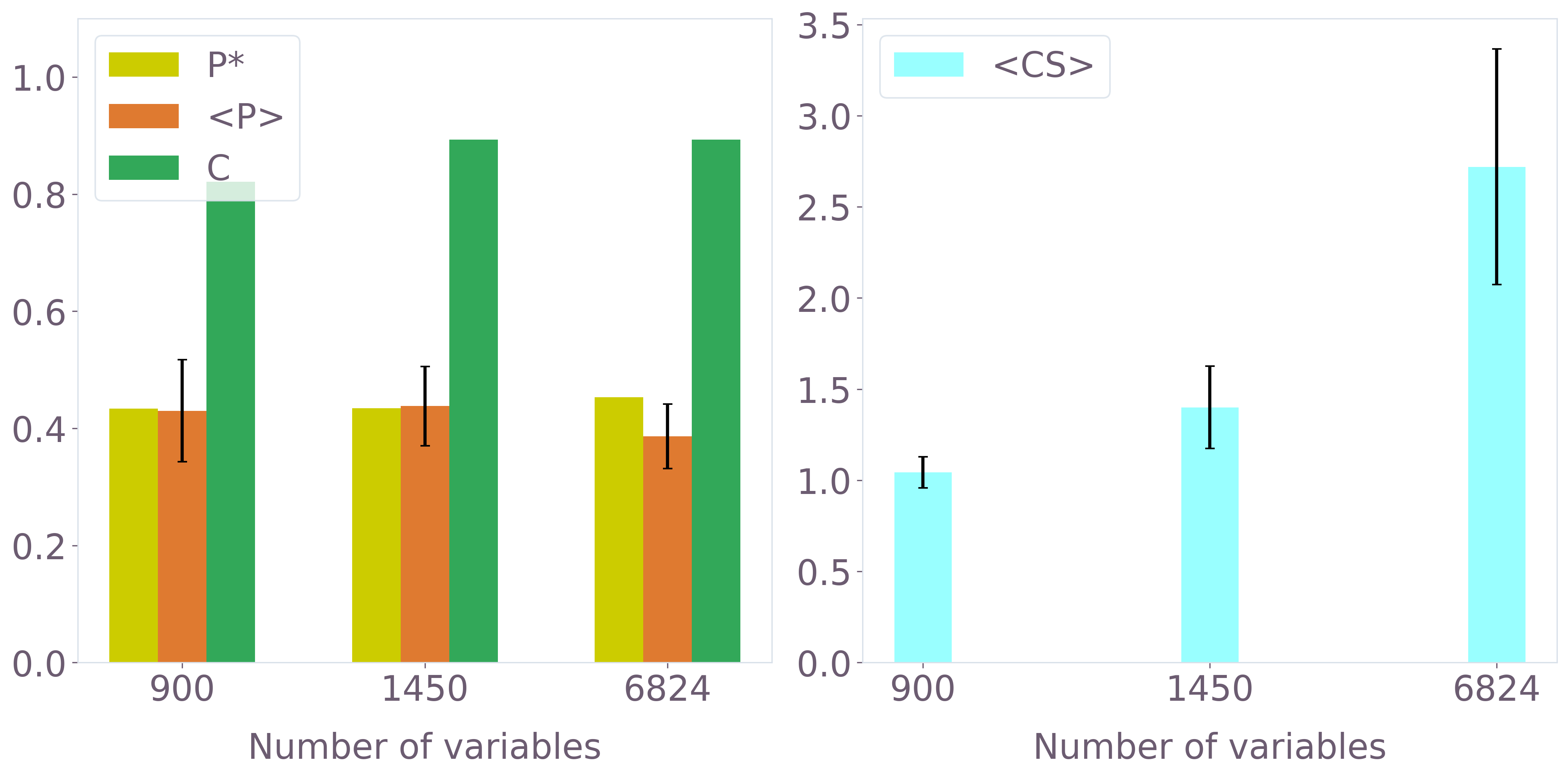}
\caption{Analysis of the hydration-sites prediction performance across different QUBO instance sizes for PDB ID $=$ 3b7e from Table~\ref{tab:test_set}, including the ligand in the 3D-RISM calculation. Top panel shows the performance analysis for the quantum optimization run using the Q-CTRL solver on the IBM devices, on the QUBO instances obtained from the following parameters: $\sigma^2 = 1.0~\textbf{\AA}^2$ and $\tau_g = 0.1$ are used uniformly, and, from smaller to larger instances, $\delta = 1.35, 1.15, 0.95~\text{\AA}$, to discretize the 3D-RISM density for the mapping onto the QUBO problem. Bottom panel shows the same analysis performed on the classical SA results on QUBO instances obtained from the following parameters: $\sigma^2 = 1.0~\textbf{\AA}^2$ and, from smaller to larger instances,  $\tau_g = 0.1, 0.05, 0.002$ and $\delta = 0.5, 0.5, 0.35~\text{\AA}$. For each panel, we report: on the left-side plot, P* (closest water placement precision), $<$P$>$ (cluster-averaged precision) and C (fraction of crystal waters identified), and on the right-side of the plot $<$CS$>$ (average cluster size); error bars show the 95\% confidence interval for $<$P$>$ and $<$CS$>$. Each metric is computed as described in the Methods section, extracting the PWs corresponding to the best solution obtained from the corresponding optimization }
    \label{fig:QS_1}
\end{figure}

Based on these considerations, we explored different mappings between QUBO problem instances and the underlying $g(\mathbf{r})$ density. We varied the parameters used to generate system ``g'' in Table~\ref{tab:systems} to give a fair estimation of the impact of these approximations on the method's performance. To do so, we reduced the $\delta$ and $\tau_g$ parameters used in our original set (see Table~\ref{tab:systems}) to increase the size and the overall complexity of the optimization problem. We ensure that the sizes of these problems are still tractable on quantum hardware. The results are displayed in the top panel of Figure~\ref{fig:QS_1}, where the performance metrics are computed extracting the best solutions obtained for each new QUBO instance using Q-CTRL's optimization solver.  
As with the cases presented in Table~\ref{tab:systems}, the quantum optimizer was also able to find the optimal solution.

Both precision metrics (P$^*$, $\langle \mathrm{P} \rangle$) and the fraction of successfully identified CWs (C) consistently increase with the size and complexity of the QUBO. The larger QUBO instance (123 variables) is able to recover up to $\sim$60\% of CWs, showing a notable increase with respect to the 71 variable instance. The metric $\langle$CS$\rangle$ exceeds the value of $1.0$, which could be related to diffused regions of high enough density which can be mapped by more than one variable. This is not possible, or less likely to happen, for smaller QUBO instances.

\begin{figure}[!htb]
    \centering
\includegraphics[width=.75\textwidth]{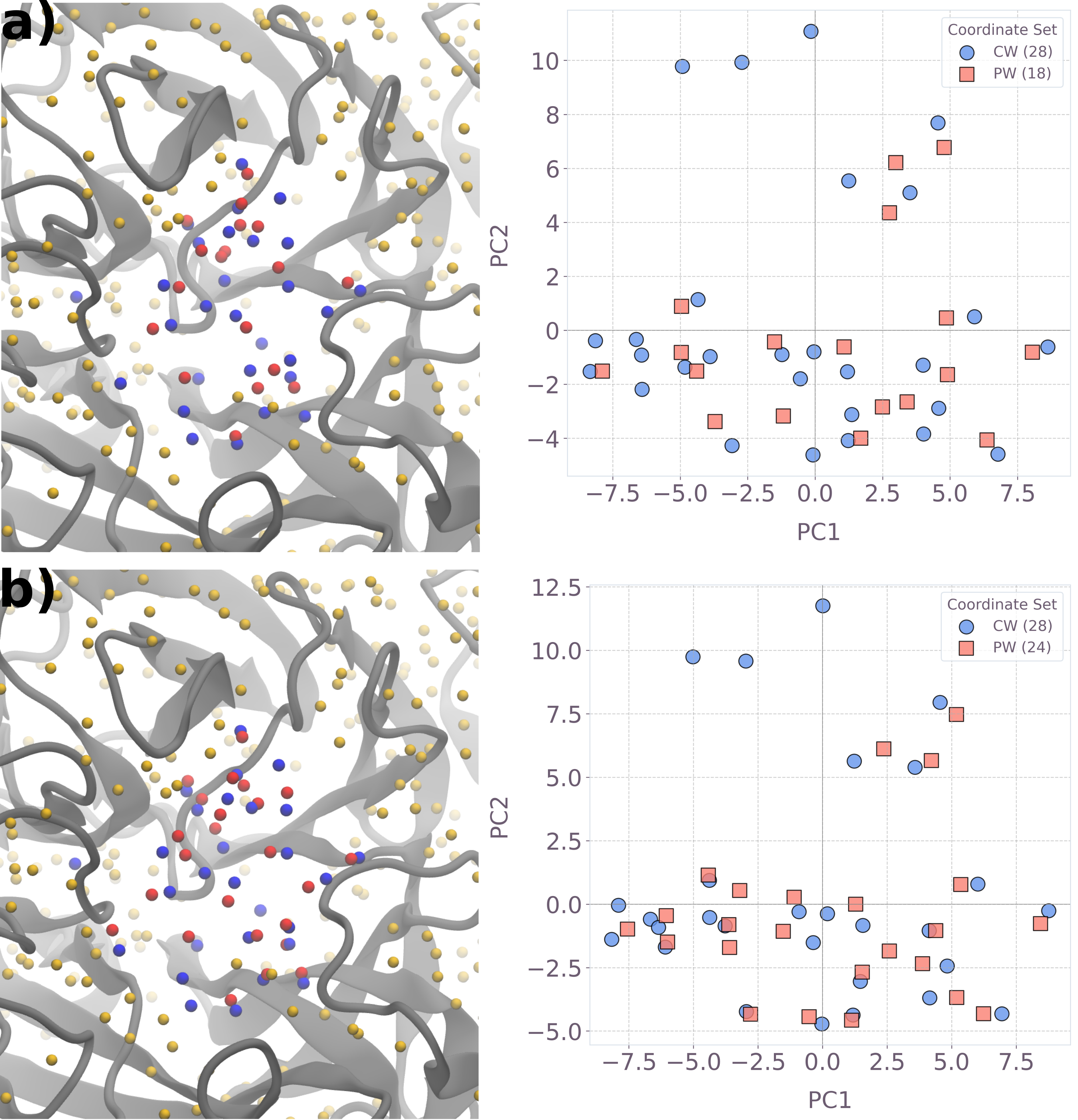}
\caption{Visualization of hydration-sites location corresponding to the best solutions obtained from: 123 (panel a) and 900 (panel b) variables instance analysed in Figure~\ref{fig:QS_1}. On the left side of each panel we show a 3D visualization of the molecular system, using VMD software; red spheres represent the oxygen atoms placed as a solution of the QUBO, blue the reference CWs, while in yellow are represented all the crystal waters available in the PDB original file. On the right-side we have a 2D space representation of the hydration-sites location obtained through a principal component analysis (PCA) performed for an easier visualization of the corresponding 3D space placement. Light-red squares represent the PW location on the principal components (PC), while the light-blue circles those of the CWs, with their respective numbers reported in the legend.}
\label{fig:WP_1}
\end{figure}

To give a better idea of the results obtained from these predictions, we report in Figure~\ref{fig:WP_1}, panel a, the 3D geometry of the identified hydration site from the best solution obtained on the 123 variable QUBO displayed on Figure~\ref{fig:QS_1}. On the left-side of the figure, the 3D structure of the binding pocket is represented in gray, together with the oxygens of CWs present in the original PDB (yellow spheres). We highlight the set of water we targeted for the hydration site prediction task in blue, while the red spheres represent the PW locations. 

The 3D visualization being hard to interpret, we performed a principal component analysis (PCA) as a dimensionality reduction technique, to better apprehend the variance (difference and spread) between the two sets of 3D coordinates, the CWs coordinates and the PWs coordinates, in a simple 2D plot. The PCA model is fitted to the combined 3D data (\textit{i.e.}, the CW and PW coordinates set combined) to find the two directions (principal components PC1 and PC2) in the 3D space that capture the maximum amount of variance between the two sets of coordinates. Both data points are then projected onto the new 2D space defined by PC1 and PC2 as shown in Figure~\ref{fig:WP_1}. As observed from the metrics analysis, not all CWs (light-blue dots) have been identified by at least one PW (light-red squares). The QUBO-based approach places roughly 60\% of the CWs and the closest PW are placed nearby the corresponding identified CW.

\subsection*{Problem scale-up}

In the previous section we showed that the discretization step and dimensionality reduction procedure we performed to map the 3D-RISM density into QUBO (see Figure~\ref{fig:flow}), has an impact on method's prediction accuracy. Therefore, to assess the full potential of our strategy, we further scaled the QUBO instances complexity and size up to better map the underlying 3D-RISM density $g(\mathbf{r)}$, and we evaluated their performances. 

We kept using the 3D-RISM density computed on PDB ID $=$ 3b7e with ligand as example while further reducing $\delta$ and $\tau_g$. We generated QUBO instances exceeding the hardware limitation of 156 binary variables (each binary variable being mapped onto 1 qubit). To perform this analysis, we had to rely on the classical SA heuristic approach to obtain the hydration-sites positions. The results are reported in Figure~\ref{fig:QS_1}, bottom panel, together with all the parameters used to generate the QUBO instances.

Compared to the 123 variables instance,  P*, <P> and C, all show a notable increase starting from the 900 variables instance. No remarkable improvements are observed beyond 900 variables on P* and <P>, while C reaches a plateau value $\sim0.9$ starting at 1450 variables. $<$CS$>$, shows a significant increase for the 6824 variables instance. Nevertheless the 55-7 fold increase in the number of variables with respect to the 123 -123 variables instance, only corresponds to a $\sim2.5$ increase in <CS>.

$<$P$>$ and $<$CS$>$ both exhibit an associated statistical variance, which we report as the 95\% confidence interval. We see how this attribute consistently decreases with the system size. We observe the same general trend in the additional results reported in Section~S4, Figures~S5 and S6, of the SI, reinforcing this evidence. This trend is not that clearly followed by <CS>. To support this observation we also report their coefficients of variation in Table~\ref{tab:CV_3b7e_L}, computed as the ratio between the standard deviation and the mean value of the computed metrics. We can conclude that the PW location with respect to the CW reference becomes gradually more robust for larger instances, since the variability of $<$P$>$ tends to become lower while its average value increases.

Before moving to the next section, we stress that the increase in the prediction accuracy and statistical consistency with the increase of the QUBO size is also observed in other examples taken from Table~\ref{tab:test_set}, varying again $\delta$ and $\tau_g$ to obtain QUBO instances of increasing complexity, spanning up to $\sim$6000 binary variables. The evaluations of the resulting hydration-sites predictions are reported in SI, Section~S4, from Figure~S3 to Figure~S6.

\subsection*{Comparison to other methods for hydration-site prediction}

\begin{figure}
    \centering
\includegraphics[width=1.0\textwidth]{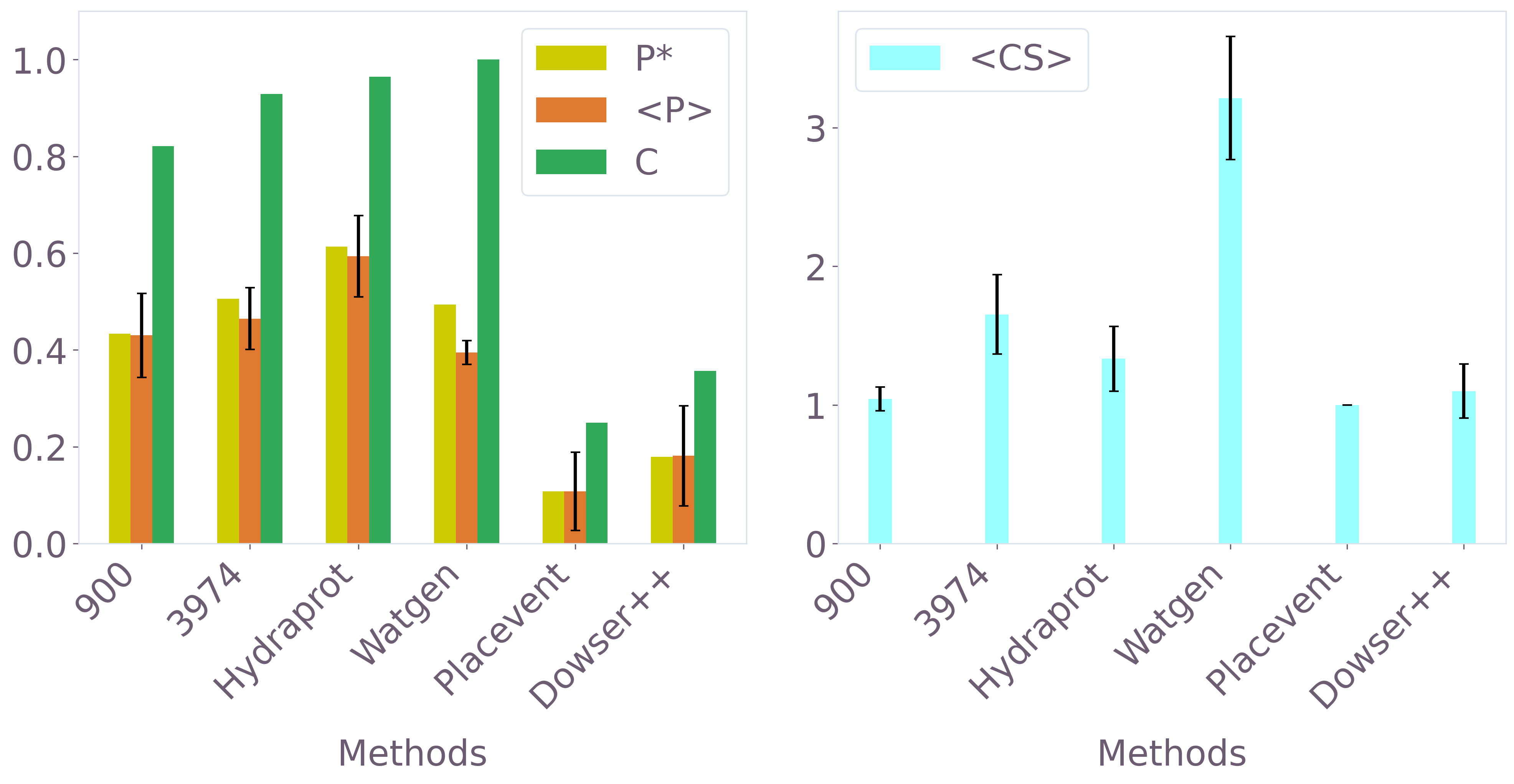}
\caption{Comparative performance analysis of different classical methods: the two best performing QUBO models 
for PDB ID $=$ 3b7e from Table~\ref{tab:test_set}, with ligand included in the 3D-RISM density, characterized by 900 variables, $\sigma^2 =1~\textbf{\AA}^2$, $\tau_g = 0.10$, $\delta = 0.5~\textbf{\AA}$, and 3974 variables with $\sigma^2 = 0.8~\textbf{\AA}^2$, $\tau_g = 0.05$ and $\delta = 0.35~\textbf{\AA}$, Hydraprot, Watgen, Placevent and Dowser++. The figure presents four performance metrics as a function of the number of binary variables in the QUBO formulation. P* (closest water placement precision), $<$P$>$ (cluster-averaged precision), C (fraction of crystal waters identified) are presented on the left plot; $<$CS$>$ (average cluster size) are showed on the right plot. Error bars show the 95\% confidence interval for $<$P$>$ and $<$CS$>$.}
    \label{fig:QS_2}
\end{figure}

The task of predicting hydration sites in proteins is the focus of a variety of methods~\cite{WM1,WM3,WM4,WP1}. In this section we compare our QUBO-based approach to a selected set of methods from the literature. Some of them are designed to be computationally affordable and can be used on modest size hardware~\cite{WP1,WM2,WM3}, at the cost of a lower accuracy in the prediction. Molecular dynamics and Monte Carlo based simulations are in comparison more demanding in computational resources, reaching higher precision in the predictions while offering a potentially complete view on the thermodynamics of the system~\cite{Blazhynska2025.10.17.683050,WM0}. In Figure~\ref{fig:QS_2}, we compare our method's performances to a limited selection of popular, fast classical methods including: Hydraprot~\cite{hydraprot}, a neural network crafted for this task, Placevent~\cite{Placevent}, Watgen~\cite{watgen} and Dowser++~\cite{dowser++}. These methods are based on different algorithms and perform therefore different types of computations with respect to each other. A short description of these methods is available in SI, Section~S3. They have been classified in the literature and range from the so-called knowledge-based to deep-learning based approaches~\cite{WM3, WM4}. They all exhibit low computational costs, only requiring, on average, just a few minutes of computing on a regular CPU. 
The selection we propose does not represent a full benchmark of the available classical methods, which is out of the scope of the present work. Among these methods, only Placevent uses 3D-RISM densities to identify the hydration sites but is important to note that Placevent does not use a QUBO formulation, as the one proposed in this work.

To perform this comparison, we selected the best performing QUBO instances for the test case of PDB ID $=$ 3b7e with ligand, discussed in more details in the SI, Figure~S5. The solutions for these instances are obtained from simulations using SA heuristic, as already explained above. All parameters used to generate the QUBO instances are reported in the caption of Figure~\ref{fig:QS_2}, where we included the results from both the 900- and the 3974-variable instances. Again, we observe an increase in the performance of the QUBO-based method passing from 900 to 3974 variables. Most notably, P and C sensibly increase, the variance on <P> is reduced even if <CS> is in average $\sim$50\% larger for the 3974 instance. In other words, with a $\sim$4.5 fold increase in the number of variables, only a $\sim$1.5 fold average increase of PW is observed.

To produce the hydration-site predictions of the selected methods from the literature, we used publicly available codes to run simulations on the same PDB ID and performed comparable performance analysis. From Figure~\ref{fig:QS_2} we observe that our QUBO approach and Hydraprot show the best performances for this test case: the precision metrics are similar between these two methods whereas they perform substantially better than the others. Watgen identifies all the CW positions, but the average cluster size <CS> is more than double, and presents a larger variance, compared to the ones of QUBO and Hydraprot. This means that Watgen has the tendency to place a large number of water beside those corresponding to the CW sites.

Most notably, Placevent, which also uses 3D-RISM as input, clearly underperformed with respect to the QUBO approach, and also with respect to all other methods. These observations are reinforced by a similar analysis performed on PDB ID $=$ 4h2F without ligand, as reported in SI, Figure~S7 therein.

We conclude that the QUBO-based approach exhibits performances on par, and in some cases superior, to other classical methods available in the literature. We can expect from our previous results that a quantum optimization will deliver results in, at least, the same range of precision when quantum hardware at the required scale is available.

\subsection*{Advantage forecasting}

Projecting when quantum computing will have the potential to outperform classical approaches is nuanced. To forecast quantum advantage for hydration-site prediction, we first need to determine when our QUBO-based hydration-site prediction method achieves performance that is competitive with alternative approaches, reliably predicting crystal water positions. Secondly, we need to assess when quantum computing can be expected to outperform classical methods in solving the corresponding QUBO problem instances. As a next step, we have to estimate the quantum resources required to reach the potential threshold. Comparing these resource requirements with quantum hardware provider roadmaps, we can then get an idea of when the required resources are likely to be available. In this work, we focused specifically on IBM hardware, as our benchmarks were run on these devices and IBM provides a detailed public roadmap \cite{noauthor_ibm_nodate}.

For the QUBO-based approach to hydration-site prediction, we have seen that it becomes practically relevant at scales on the order of 1,000 variables. For the 3b7e-with-ligand case within the specified grid extent, a grid spacing of 0.5\,Å, corresponding to 900 variables (see Figure~\ref{fig:QS_1}, bottom panel), can be taken as an indicative threshold for the onset of competitiveness with alternative approaches (see Evaluation of hydration-site predictions section). Consequently, the ability to reliably solve instances at this scale establishes a lower bound for the definition of a potential quantum advantage threshold. 

Achieving quantum advantage requires outperforming any classical algorithm, whether exact or heuristic. An important milestone on the path towards full quantum advantage is the ability to outperform exact classical solvers, a regime referred to as quantum utility by IBM. As we have seen in our QPU benchmarks, an exact classical solver already struggles to solve the QUBO problem instances at a size of 123 variables, so the 900-variable instance lies well within this regime and can therefore be retained as a reasonable potential quantum utility threshold.

\begin{figure}
    \centering
\includegraphics[width=0.5\linewidth]{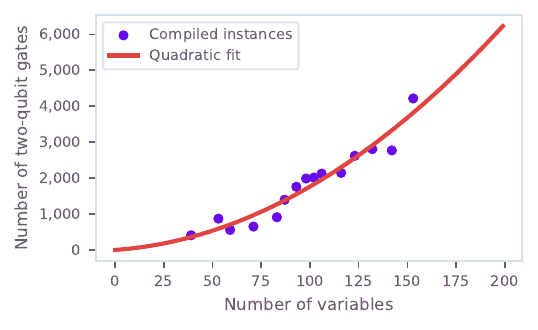}
\caption{Two-qubit gate scaling with problem size. For the 3b7e-with-ligand structure, purple dots show two-qubit gate counts after compilation for varying numbers of variables determined by the grid spacing. A quadratic fit (red line) is used for extrapolation.}
    \label{fig:resource_scaling}
\end{figure}

To estimate quantum resource requirements at this potential quantum utility threshold, we compiled instances up to 156 variables to a 156-qubit Heron device and extrapolated the scaling (Figure~\ref{fig:resource_scaling}). The two-qubit gate count scales approximately quadratically with the number of variables; under this scaling, a 900-variable instance would require on the order of 100,000 two-qubit gates. Comparing with IBM's quantum development roadmap \cite{noauthor_ibm_nodate}, qubit counts around 1,000 will likely be available by 2027. The projected two-qubit gate requirements, however, remain out of near-term reach and will likely require early forms of error correction, projected for 2029. 
Notably, the higher (square) connectivity in next-generation hardware, along with improved compilation techniques, may reduce the number of gates required, making the target problem scales plausibly attainable as early as 2028.

To achieve full quantum advantage, the quantum approach must ultimately outperform not only exact but also heuristic classical solvers. However, a quantitative comparison with classical heuristics, such as SA, at the target scale is not yet possible because the required quantum resources are not available. Algorithmic considerations and empirical studies suggest that SA will increasingly struggle to find high-quality solutions within practical runtimes as problem size grows \cite{johnson_optimization_1989, johnson_optimization_1991, henderson_theory_2003}, whereas the quantum solver may benefit from a more structured exploration of the search space, albeit subject to potential expressibility limitations in the low-depth limit \cite{streif_comparison_2019, zhou_quantum_2018}. In line with the benchmarking results presented in the QPU benchmarks section, we therefore regard it as plausible that the Q-CTRL solver will achieve performance at least comparable to SA once hardware is sufficiently stable to support reliable circuit execution at the target scale of 900 variables. However, ultimately, any claim of quantum advantage must be supported by empirical evidence and comparisons with state-of-the-art classical approximate solvers; based on the above resource estimation and IBM’s hardware roadmap, we anticipate that the necessary tests may become feasible by 2028.

\section*{Discussion}

We tested our QUBO-based method to identify 3D hydration sites in protein pockets, using a quantum solver to find the corresponding optimal solution. Hardware calculations have been run on two Heron IBM's devices (R2 and R3) up to 123 qubits and have been complemented by larger classical QUBO simulations to estimate the accuracy of the approach. Various protein-ligand complexes of interest for health care have been used as test cases, demonstrating the feasibility of the simulations on NISQ devices, using Q-CTRL's optimization solver including Fire Opal automated error suppression~\cite{ES_1,ES_2}.

Comparing the small instances attainable on quantum hardware ($\sim$100 variables) to larger instances that can be approximately solved through classical QUBO simulations ($\sim$1000 variables) we showed how the method's accuracy scales toward improved solution qualities. For most cases, a good correspondence between the predicted water placement and the reference experimental measures is achieved. We also compared our method to a representative set of other classical approaches known in the literature, that are not using the QUBO formulation we proposed. From this comparison, we observed that our QUBO approach performs on par with the best fast hydration-sites prediction methods we could test.

Moreover, we showed that, even around $\sim$100 variables, classical optimization solvers fail to provide a provably optimal solution within a reasonable computing time, whereas quantum optimizations or classical heuristics outperform their best found approximate solution. To further estimate the quantum resource required at the utility threshold, we performed a detailed resource analysis based on number of optimization variables and on two-qubit gate counts. A 900-variable instance (900 qubits), which would allow us to tackle most of the industry test cases, requires on the order of 100,000 two-qubit gates and some early forms of error correction. This should make the target problem reachable within 5 years.

This opens to the possibility that quantum optimization might deliver a quantum advantage by providing better solutions, or equally good solutions but at a lower computational cost. It also demonstrates how practical quantum computing could rapidly integrate CADD pipelines to perform meaningful computations, as in the step needed to prepare protein-ligand complexes for MD simulations required for drug lead optimization or docking calculations.

\section*{Methods}

In this section, we explain step-by-step our hydration-sites prediction workflow: from approximating the input 3D-RISM density as a Gaussian Mixture Model (GMM), to optimizing the GMM using a QUBO formalism, to the implementation of the method on the quantum hardware using Q-CTRL's optimization solver on real hardware. 

\subsection*{Step-by-step workflow for the QUBO-based 3D hydration-site prediction}
\label{TSSIP}

\begin{figure}[!ht]
\begin{center}
\includegraphics[width=0.75\textwidth]{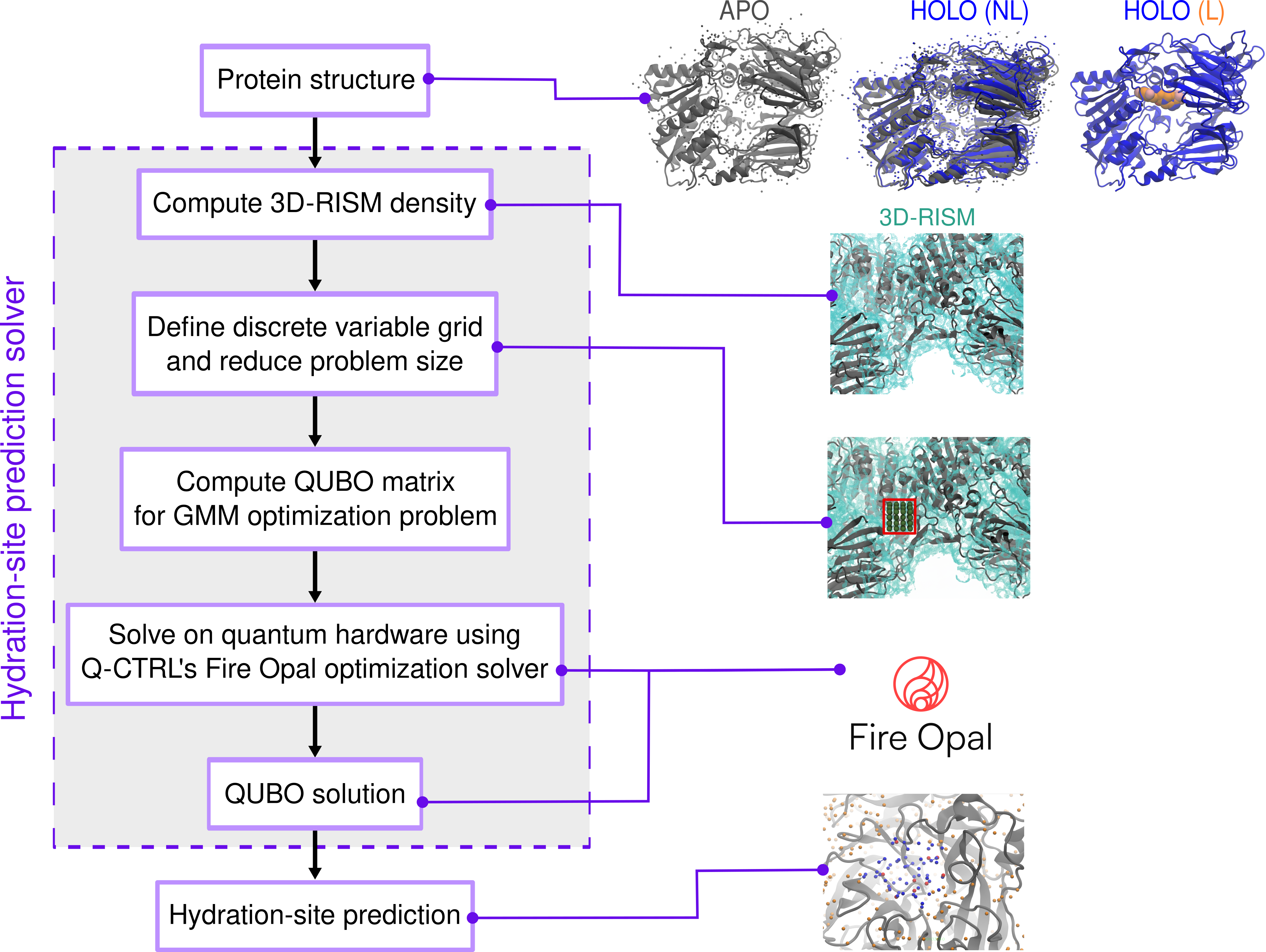}
\end{center}
    \caption{Workflow illustrating the full-stack conversion of the hydration prediction problem into its Quadratic Unconstrained Binary Optimization (QUBO) form, followed by the translation to a QAOA-like hybrid quantum algorithm, from which we predict the hydration sites in a protein. An initial 3D-RISM continuous water density $g(\mathbf{r})$ within the protein of choice is converted into a binary variable placement grid. This grid is used to evaluate the QUBO matrix elements that map the information contained in $g(\mathbf{r})$ into the combinatorial optimization form that we use to find the GMM that best approximates the target density $g(\mathbf{r})$. The QUBO matrix can now be injected into the digital quantum optimization solver including the error suppression pipeline. The output directly reads as coordinates in 3D space of the most likely sites for stable water molecules. Example structures of an apo (gray), holo without ligand (blue) and holo with ligand (ligand in orange) are added to the diagram. A 3D-RISM density isosurface is showed in cyan, as well as a grid of variable generated inside a focused region (green spheres inside a red square)}
    \label{fig:flow}
\end{figure}

In Figure~\ref{fig:flow} we show a diagram with the workflow we followed to integrate the QUBO formulation in a computational pipeline to perform the hydration-sites prediction simulations. It represents the sequence of computational tasks we apply to perform the simulations described in the Results section, and to evaluate their accuracy. 

The overall procedure requires to first select the 3D molecular structure that constitute the target of the hydration site prediction. This structure is needed to compute the 3D-RISM density, that bears the essential information about the water distribution around the solvated molecule, a protein in our case. A restricted area of the molecular structure is selected, where our attention is focused to perform the hydration-sites prediction. A set of discrete variables is used to map the localized area's 3D-RISM density onto a QUBO problem, resampling such underlying 3D-RISM density on a coarse grid to reduce the problem size and complexity. Using this variable grid, the problem is cast into a GMM optimization problem, and the corresponding QUBO matrix is computed. Q-CTRL's Fire Opal solver is used to solve the optimization problem on a quantum hardware, and the best solution obtained is interpreted as the set of positions of the predicted water molecules.

\subsubsection*{3D-RISM density computations}

The very first step consists in selecting from Table~\ref{tab:test_set} an X-ray structure and extracting the atomic coordinates of the system to compute the 3D-RISM continuous water density, as illustrated in Figure~\ref{fig:flow}. To compute the 3D-RISM continuous water density $g(\mathbf{r})$ we used the popular Ambertools package, a widely used package in the community of MD simulations~\cite{R2_3DRISM,R3_AT}. 

The 3D-RISM theory provides probability density of all the solvent molecules around a solute on a set of given positions in 3D space $\mathbf{r}$. It gives as output one density function for each atom type of the solvent molecule~\cite{R2_2_3DRISM2,R2_3DRISM,R3_AT}. In the case of water, this means that two different sets of density can be computed, one describing the distribution of oxygen atoms, and one of the hydrogen atoms. High values of $g(\mathbf{r})$ correspond to regions with high solvent concentration compared to the pure bulk solvent. One can interpret those high density regions as location of stable water molecules, due to strong water-solute interactions. In our work, the solute is represented always by proteins or proteins in a complex with a ligand.

3D-RISM requires as input the 3D coordinates of each atom of the molecular structure of the solute which one needs to solvate. In our case, the proteins molecular structures are stored in the PDB files we selected. In addition to the molecular structures, 3D-RISM also requires the force field parameters that are needed to solve the integral equations that are the very core of the method~\cite{R2_3DRISM,R2_2_3DRISM2}. For the protein, we used the ff14SB~\cite{ff14sb} Amber force field, while the popular SPC/E water model was used to describe the solvent~\cite{spce}.

For the five holo structures presented in Table~\ref{tab:test_set}, we provided two sets of 3D-RISM densities, one computed with the effect of the ligand, and one without it. In such cases, we used the same PDB ID, the protein structure itself being unchanged, while the electrostatic and steric effects of the ligand in the binding pocket were taken into account (L) or ignored (NL), resulting in a different 3D-RISM density.
This is a pure model-related information, that can be used to estimate the ligand effect on the computed 3D-RISM density, and its impact on the hydration-sites prediction. PDB ID $=$ 3beq is the only case for which we provide only one 3D-RISM density, since it represents the apo structure of PDB ID $=$ 3b7e. 
For each PDB ID, and each L and NL structure, the density function $g(\mathbf{r})$ was computed on a grid with 120x120x120 points with spacing $\delta = 0.5~\textbf{\AA}$. The only exception is represented by PDB ID $=$ 1x70, for which larger grids have been used, with 200x200x200 points for case L and 180x180x180 points for case NL, respectively. The computed densities were saved in a \textit{.dx} format, which contains the information on the geometry of the grid and on the point-by-point values of the computed functions. We ran the 3D-RISM simulations on standard CPUs, with an average simulation time of $\sim$30 minutes. For the larger boxes presented in our tests, computing times were sensibly larger, in the order of $\sim$3 hours. It is worth noting that machine learning models able to predict the initial 3D-RISM densities have been proposed which might potentially improve the method accuracy while reducing the required computational time~\cite{R4_3DRISM_ML1,R4_3DRISM_ML_2}. Moreover, it is always possible to rely on high performance computing to reduce execution time, which was not the intention in the present work.

\subsubsection*{Discrete variables grid and problem size reduction}

The original grids used for the computation of the 3D-RISM densities described above, encompass a number of points in the order of $\sim$1000000. The quantum hardware resources presently available are insufficient for this problem sizes, while classical computational strategies to solve QUBO problems of such dimensionality become extremely expensive. Therefore, following the workflow presented in Figure~\ref{fig:flow}, we operated a dimension reduction of the problem. Discretizing the $g(\mathbf{r})$ function on coarse grids, we generated QUBO problems of adequate sizes for the quantum calculations performed above. First we focused only on the binding site of the protein. This choice is motivated by the fact that the binding site is often the most relevant portion of the protein for a drug discovery project. We further reduce the size of the problem by selecting the density to be considered in the QUBO formulation. We define a density threshold $\tau_g$ to ignore locations where the density is lower than a chosen value. This step is not necessary to model the hydration-site prediction problem as a QUBO, but only required to handle the practical constrains imposed by the quantum hardware on the one hand, and to reduce the classical simulations costs on the other. 

In practice, to operate the dimensionality reduction described above, after having computed $g(\mathbf{r})$ given a protein .pdb, we proceeded as follows:
\begin{enumerate}
    \item We select a cubic subdomain of the density including the protein binding pocket; 
    \item The crystal waters present in the binding pockets are selected and they coordinates are saved, since they will be later used to evaluate the performance of the hydration-site prediction~\cite{R1_FDA}.
    \item We define a QUBO problem inside the selected region, where the variables are arranged in a regular cubic grid $\mathbb{G}$ with defined grid spacing $\delta$; 
    \item We keep only the grid points located in correspondence to, or close to, a $g(\mathbf{r})$ value $>\tau_g$; 
    \item The QUBO matrix is computed on the resulting collection of grid points, using a uniform Gaussian variance at each point, so that  $\sigma_i^2 \equiv \sigma^2  \quad \forall i \in \mathbb{G}. $
\end{enumerate}

To prepare for hardware tests, we consider cubic subdomains of 15$\text{\AA}$ sides, and use a coarser grid spacing compared to the original 3D-RISM grid setup, with QUBO grid points spacing larger than the original 0.5$\textbf{\AA}$ in the 3D-RISM density. The main parameters of the grids used are reported in Table~\ref{tab:systems}. More information about the effects of the choice of $\delta$ and $\sigma^2$ are discussed in SI, Section~S2 therein. The grid generated in this way is used to formulate the hydration-site prediction problem as a GMM optimization, cast as a QUBO, as shown in Figure~\ref{fig:flow}. The QUBO matrix elements are computed on this grid and the details of the QUBO formulation of the hydration-sites prediction problem are described in the next section.

\subsection*{QUBO formulation of the 3D hydration-site prediction problem}

While methods like the 3D Reference Interaction Site Model (3D-RISM) provide a continuous density map of water molecule distribution on a spatial grid, the explicit location of stable water molecules and their thermodynamics is a complementary valuable information. Approaches such as Placevent~\cite{Placevent} and GaSol~\cite{fusani_optimal_2018} work in this direction, extracting an optimal water network coerced by the continuous 3D-RISM distribution.

Some of the authors proposed an alternative approach to transition from the continuous density to a discrete optimization problem, approximating the 3D-RISM density distribution with a GGM~\cite{QP}. Each Gaussian component of the GMM represents a potential, localized hydration site, transforming the problem from continuous sampling to the selection of an optimal subset of discrete sites~\cite{QP}. The problem then becomes to find the GMM that best fits the given 3D-RISM density. 

To do so, we consider a set of $N$ potential hydration sites: our aim is to associate to each site a Gaussian center in the optimal GMM, that we need to determine. We introduce a set of binary variables, $\mathbf{x} = (x_1, x_2, \dots, x_N)$, where each $x_i \in \{0, 1\}$ is associated with the $i$-th Gaussian component of the GMM~\cite{QP}, with:
\begin{itemize}
    \item $x_i = 1$ if a water molecule is placed at the $i$-th potential site
    \item $x_i = 0$ if no water molecule is placed at the $i$-th site.
\end{itemize}

The central challenge is to select a configuration of water sites, $\mathbf{x}$, that minimizes the total system energy, while satisfying physical constraints, such as the minimum repulsive distance between two water molecules. In our approach all the thermodynamics information is contained in the 3D-RISM density function $g(\textbf{r}): \mathbb{R}^3 \to \mathbb{R}$, with $\textbf{r}$ being a spatial coordinate. This function, that can be computed at a reduced computational cost on a pre-defined grid of points, has large positive values in correspondence of locations where the water-water and water-protein interactions are favorable~\cite{R2_3DRISM,QP}. Therefore we aim at finding the GMM that minimizes the following L$^2$-norm~\cite{QP}

\begin{equation}\label{eq:l2distance}
    I^2 := \int_\mathcal{C} \left( g(\textbf{r}) - \sum_{i=1}^M \mathcal{G}(\textbf{q}_i, \sigma^2)(\textbf{r}) \ x_i  \right)^2 d\textbf{r}.
\end{equation}

Expanding the square and with minimal manipulation of the resulting terms, it is easy to see how this can be written as a Quadratic Unconstrained Binary Optimization (QUBO) cost function. We refer to the original work for more details on the derivation~\cite{QP}. In Eq.~\eqref{eq:l2distance},  we denote by $\mathcal{G}(\textbf{q}_i, \sigma^2)$ an isotropic Gaussian with mean $\textbf{q}_i \in \mathbb{R}^3$ and variance $\sigma^2$. The set of $\textbf{q}_i$ defines the grid in 3D space where the binary variables are placed. Note that this grid can correspond to the 3D-RISM one.

The hydration-sites prediction problem is solved by minimizing a quadratic objective function $C(\mathbf{x})$, as defined by the standard QUBO equation:
\begin{equation}
    \min_{\mathbf{x} \in \{0, 1\}^N} C(\mathbf{x}) = \sum_{i=1}^N Q_{ii} x_i + 2\sum_{i<j}^N Q_{ij} x_i x_j.
    \label{eq:cost_function}
\end{equation}
In this context, the coefficients of the symmetric QUBO matrix $\mathbf{Q}\in \mathbb{R}^{N,N}$ are specifically engineered to encode the energy landscape of water placement:
\begin{enumerate}
    \item The linear terms $Q_{ii}$ correspond to the linear bias applied to the binary variable $x_i$. This term can be seen as a self-energy term, connected to the solvation free energy of placing a single water molecule at site $i$, reflecting its interaction with the static protein environment (e.g., protein residues and bulk solvent). A negative $Q_{ii}$ value indicates an energetically favorable site.
    \item Quadratic terms $Q_{ij}$ explicitly encode the pairwise interaction energy between two Gaussians, placed at sites $i$ and $j$. This term is critical for introducing constraints. For two sites $i$ and $j$ that are too close, for example, $Q_{ij}$ becomes a large positive penalty value, pushing the solver towards configurations that avoid too close water molecules.
\end{enumerate}

By minimizing $C(\mathbf{x})$, a QUBO solver identifies the optimal configuration of water sites $\mathbf{x}^*$ that achieves the lowest overall energy, while simultaneously satisfying the necessary spatial separation constraints via the penalty terms~\cite{QP}. More details on the parameters needed to evaluate the QUBO matrix elements are discussed in SI in Section~S1 and~S2.

In next section, we describe how we use the quantum hardware, in combination to Q-CTRL's Fire Opal, to perform the optimization needed to perform the hydration-site prediction.

\subsection*{Hardware deployment of variational quantum optimization via Q-CTRL's Fire Opal}
Solving binary combinatorial optimization problems, such as the QUBO instances defined for the hydration-site prediction, is a challenging task for classical computers. Exact methods scale poorly, motivating heuristics that seek high-quality solutions within practical runtimes. While fully fault-tolerant quantum computing remains out of reach on today's noisy hardware, variational quantum algorithms have emerged as practical quantum heuristics with potential benefits over classical alternatives. In particular, variational quantum optimization partitions computation between quantum and classical resources: a quantum computer is used to sample from a parameterized quantum circuit, while a classical routine iteratively updates the circuit parameters to minimize an objective function encoding the target problem.

The prototypical variational quantum optimization algorithm is the Quantum Approximate Optimization Algorithm (QAOA), which applies layers of alternating cost and mixer unitaries, implementing the problem’s cost function and facilitating transitions between states, respectively, to an equal superposition initial state \cite{farhi_quantum_2014}. While QAOA admits theoretical performance guarantees in the large-depth limit of many layers for certain problem types \cite{noauthor_evidence_nodate, boulebnane_solving_2022, omanakuttan_threshold_2025}, naive, small-depth implementations on noisy hardware often fail to deliver high-quality solutions at practically relevant scales due to compilation overhead, device noise, and challenging optimization landscapes \cite{wang_noise-induced_2021, weidenfeller_scaling_2022, pelofske_short-depth_2024}. Realizing benefits therefore requires an end-to-end pipeline with system-level optimization across all stages.

Thus, to be able to reliably tackle the hydration-site prediction QUBO problem on IBM quantum devices, we employ Q-CTRL's Fire Opal Optimization Solver. As detailed in reference \cite{ES_2}, the workflow comprises of:

\begin{itemize}
\item Ansatz setup: The input problem is specified as a cost polynomial
\begin{equation}
C(\mathbf{x}) = \mathbf{x}^T \mathbf{Q} \mathbf{x} =  \sum_{i=1}^N Q_{ii} x_i + 2 \sum_{i<j}^N Q_{ij} x_i x_j, \quad \mathbf{Q} = \mathbf{Q}^T \in \mathbb{R}^{N,N} \text{, }  \mathbf{x} \in \{0,1\}^N
\end{equation}
generated from the QUBO problem matrix $\mathbf{Q}$ as introduced in Eq.~\eqref{eq:cost_function}. This objective is used to set up an enhanced variational ansatz circuit compared to standard QAOA. Note that while for the hydration-site prediction we focus on quadratic cost functions, the solver also natively supports higher-order cost polynomials.

\item Circuit compilation: To minimize overhead in the hybrid loop, the ansatz circuit is subjected to efficient parametric pre-compilation, making use of an enriched gate set with pre-calibrated, pulse-level instructions.

\item Error-suppressed hardware execution: Within the solver workflow, Q-CTRL's performance management software applies an automated error-suppression pipeline that deterministically reduces the impact of noise and realizes maximum hardware performance without additional sampling overhead \cite{ES_1}.

\item Classical parameter optimization: The circuit parameters are updated by a classical optimizer configured to maximize the efficiency of quantum-resource utilization.

\item Classical postprocessing: Measured bitstrings are refined via efficient local search correcting for uncorrelated bitflip errors.
\end{itemize}

Leveraging this combination of techniques, the solver has demonstrated that nontrivial binary combinatorial optimization problems can be solved at full device scale, outperforming competing approaches on published benchmarking instances \cite{ES_2}, and enabling record-scale solutions to real-world optimization problems \cite{noauthor_accelerating_2025}.

\subsection*{Performance metrics}
\label{sec:metric_cw}

Using the combination of the technologies described above, we were able to extract from the optimal solution obtained on the quantum hardware the discrete positions of the optimal network of water molecules. We assess the quality of the predicted water (PW) positions against experimentally determined crystal waters (CWs), which are treated as true hydration sites \cite{WM1,WM2,WM3,R1_FDA}. 

Let $n$ be the total number of CWs and the set $\{a_i\}_{i=1}^{n}$ with $a_i \in \mathbb{R}^{3}$ their positions in 3D space. We consider that the QUBO based method is able to place $m$ PWs, with $\{b_j\}_{j=1}^{m}$ with $b_j \in \mathbb{R}^{3}$ the set of their positions. $m$ can either be greater, smaller or equal to $n$. We evaluate if at least one of the PWs matches one of the reference CWs. Each $a_i \in \mathbb{R}^{3}$ is considered as a center of a sphere of radius $R_s = $3~\textbf{\AA} and if at least one of the $b_j \in \mathbb{R}^{3}$ is found, the CW is considered as successfully identified. 

The number $n*$ of PWs that successfully identified CWs, is use to compute the fraction of identified hydration sites
\begin{equation}
    C = \frac{n*}{n} \in [0,1],
\end{equation}
with $C$ close to 1 meaning that the PWs successfully cover the set of reference crystal waters.
The threshold of 3~\textbf{\AA} has been chosen to mimic a medium-low crystal structure resolution, being 1.5~\textbf{\AA} or below generally considered a good resolution.

We complement this information by quantifying how many PWs have been found in the 3~\textbf{\AA} radius spheres around each CW, and their position. 
Let $\mathcal{C}_i$ be the PWs cluster around the successfully identified CW $i$, and $N_i = |\mathcal{C}_i|$ be its size.
We use the union of all the $\mathcal{C}_i$ clusters to evaluate additional metrics as 

\begin{equation}
    \label{eq:p_average}
    \langle P \rangle = \frac{1}{n} \sum_{i=1}^{n} \left( \frac{\langle \sum_{b_j \in \mathcal{C}_i} S_{i,j}\rangle}{n} \right), \quad \text{with} \quad S_{i, j} = \frac{1}{1 + \norm{a_{i} - b_{j}}} \in [\frac{1}{1+R_s},1],
\end{equation}

and 

\begin{equation}
    \label{eq:cs}
    \langle CS \rangle = \frac{\sum_{i=1}^{m} N_i }{m},
\end{equation}
the averaged precision (Eq.~\eqref{eq:p_average}) and the average cluster size (Eq.~\eqref{eq:cs}), respectively. 
$\langle P \rangle$ scores the quality of identified hydration sites based on both the number of PWs that identify a CW and the inverse of the Euclidean distance between them and the CWs.
$S_{i, j}$ by itself accounts for the distance effect, rewarding closer placements. $\langle P \rangle$ penalizes placement methods that provide large, diffuse clusters, overestimating the number of PWs around a single CW position. At the same time, it penalizes methods that are not able to successfully identify all CWs positions. $\langle CS \rangle$ gives a quantitative information on average size of the identified clusters $\mathcal{C}_i$.
Both metrics are designed so that a value close to 1 represent a good scoring for an hydration sites identification method.
For both $\langle P \rangle$ and $\langle CS \rangle$  we evaluate the $95\%$ Confidence Interval ($\text{CI}_{95}$), computed using the standard error of the mean from their standard deviation.

In addition to their average and  $\text{CI}_{95}$, we compute the coefficient of variation ($\text{CV}$), that measures relative variability of the metrics, and is defined as the ratio of standard deviation and the mean. A robust prediction is associated to a low CV value. 

At last, we complement these metrics with the single closest water precision ($P^*$), defined as

\begin{equation}
    \label{eq:p_single}
    P^{*} = \frac{1}{n} \sum_{i=1}^{n} S^{*}_{i} \quad \text{with} \quad S^{*}_{i} = \max_{b_j \in \mathcal{C}_i} \left( S_{i, j} \right).
\end{equation}
This metric quantifies how precise is the identification of each hydration sites CW based on the single closest PW position, $b^*_{i}$, within the cluster $\mathcal{C}_i$, ignoring the cluster size and the other points of the cluster.
\section*{Acknowledgment}

This work was funded in part by the European Research Council (ERC) under the European Union's Horizon 2020 research and innovation program (grant agreement No 810367), project EMC2 (JPP). 
We acknowledge the use of IBM Quantum services for this work. The views expressed are those of the authors, and do not reflect the official policy or position of IBM or the IBM Quantum team.

\section*{Competing interests}
JPP is shareholder and co-founder of Qubit Pharmaceuticals. All the remaining authors declare no conflict of interest.

\printbibliography
\end{document}